\documentclass[conference]{IEEEtran}
\IEEEoverridecommandlockouts
\usepackage{cite}
\usepackage{amsmath,amssymb,amsfonts}
\usepackage{algorithmic}
\usepackage{graphicx}
\usepackage{textcomp}
\usepackage{xcolor}
\usepackage{subfigure}

\def\BibTeX{{\rm B\kern-.05em{\sc i\kern-.025em b}\kern-.08em
    T\kern-.1667em\lower.7ex\hbox{E}\kern-.125emX}}
\begin{document}

\title{Affine Frequency Division Multiplexing With Index Modulation}\vspace{-4mm}
\author{\IEEEauthorblockN{Yiwei~Tao\IEEEauthorrefmark{1},~Miaowen~Wen\IEEEauthorrefmark{1}, Yao~Ge\IEEEauthorrefmark{2}, and Jun Li\IEEEauthorrefmark{3}
}
	\IEEEauthorblockA{\IEEEauthorrefmark{1}School of Electronic and Information Engineering, South China University of Technology, Guangzhou 510641, China}
	\IEEEauthorblockA{\IEEEauthorrefmark{2}Continental-NTU Corporate Lab, Nanyang Technological University, 639798, Singapore}
	\IEEEauthorblockA{\IEEEauthorrefmark{3}Research Center of Intelligent Communication Engineering,\\ School of Electronics and Communication Engineering, Guangzhou University, Guangzhou 510006, China}
		Email: eeyiweitao@mail.scut.edu.cn, eemwwen@scut.edu.cn, yao.ge@ntu.edu.sg, lijun52018@gzhu.edu.cn
\vspace{-0.5cm}}


\maketitle

\thispagestyle{empty}
\pagestyle{empty}


\begin{abstract}
Affine frequency division multiplexing (AFDM) is a new multicarrier technique based on chirp signals tailored for high-mobility communications, which can achieve full diversity.
In this paper, we propose an index modulation (IM) scheme based on the framework of AFDM systems, named {\em AFDM-IM}.
In the proposed AFDM-IM scheme, the information bits are carried by the activation state of the subsymbols in discrete affine Fourier (DAF) domain in addition to the conventional constellation symbols.
To efficiently perform IM, we divide the subsymbols in DAF domain into several groups and consider both the localized and distributed strategies.
An asymptotically tight upper bound on the average bit error rate (BER) of the maximum-likelihood detection in the existence of channel estimation errors is derived in closed-form.
Computer simulations are carried out to evaluate the performance of the proposed AFDM-IM scheme, whose results corroborate its superiority
over the benchmark schemes
in the linear time-varying channels.
We also evaluate the BER performance of the index and modulated bits for the AFDM-IM scheme with and without satisfying the full diversity condition of AFDM.
The results show that the index bits have a stronger diversity protection than
the modulated bits even when the full diversity condition of AFDM is not satisfied.

\end{abstract}

\vspace{-2mm}
\section{Introduction}
The next generation of communication standards puts forward a new vision for the advancement of wireless communication technologies.
Achieving ultra-high reliable communication in high mobility scenarios such as autonomous driving and high-speed railway scenarios has become an important topic~\cite{10152009}. 
In this scenario, the wireless channel has a large Doppler shift and can be modeled as a linear time-varying (LTV) channel.
Orthogonal frequency division multiplexing (OFDM) as a classical wireless communication technique has been widely used in current communication systems~\cite{4607239}. However, the orthogonality between subcarriers of OFDM systems is destroyed over the LTV channels, and the resulting inter-carrier interference (ICI) deteriorates the performance of OFDM systems.

So far, many variants of communication systems have been conceived to cope with the high-mobility scenarios.
A precoding and detection scheme for the OFDM system has been proposed in~\cite{7027287}.
This scheme divides the transmission symbols into several subblocks and inserts redundant symbols between the subblocks to eliminate the ICI.
An orthogonal chirp division multiplexing (OCDM) scheme has been proposed in~\cite{7523229}, which exploits multipath diversity to achieve better performance than the OFDM scheme.
Besides, an orthogonal time-frequency space multiplexing technique (OTFS) has been developed in~\cite{7925924}.
The OTFS scheme modulates the transmitted symbols over the delay-doppler domain, and multiplexes each transmitted symbol into the frequency-time domain by using the inverse symplectic finite Fourier (ISFFT) transform.
In this manner, the transmitted symbols pass through an equivalent time-invariant channel, thus overcoming the interference of the LTV channel.
The OTFS scheme can achieve excellent performance gains compared to the OFDM scheme over LTV channels. However, the two-dimensional structure of the OTFS results in a high pilot overhead due to extended transmission resources occupied by pilot symbols.

Recently, an affine frequency division multiplexing (AFDM) technique has been proposed in~\cite{9473655} and~\cite{10087310}, which is based on the discrete affine Fourier transform (DAFT).
The experiments have proved that the AFDM scheme achieves the same performance as the OTFS scheme over the LTV channel.
{Besides, with specific parameter settings, the AFDM scheme can achieve full diversity.}
This scheme modulates the transmitted symbols in the discrete affine Fourier (DAF) domain and then transforms the transmitted symbols to the time domain by using inverse DAFT.
{Since it requires only one dimension of transformation, the AFDM system is less complex to implement than the OTFS system.}
Attracted by its advantages, researchers have begun to explore improvements and applications of the AFDM scheme. For instance, AFDM-based integrated sensing and communications have been proposed in~\cite{9940346}, which have high sensing accuracy in high mobility scenarios.
For the channel estimation of the AFDM scheme, a single-pilot and a multi-pilot-assisted channel estimation approaches have been presented in~\cite{9880774}.

\begin{figure*}[t]
	\center
	\includegraphics[width=7.0in,height=1.3in]{{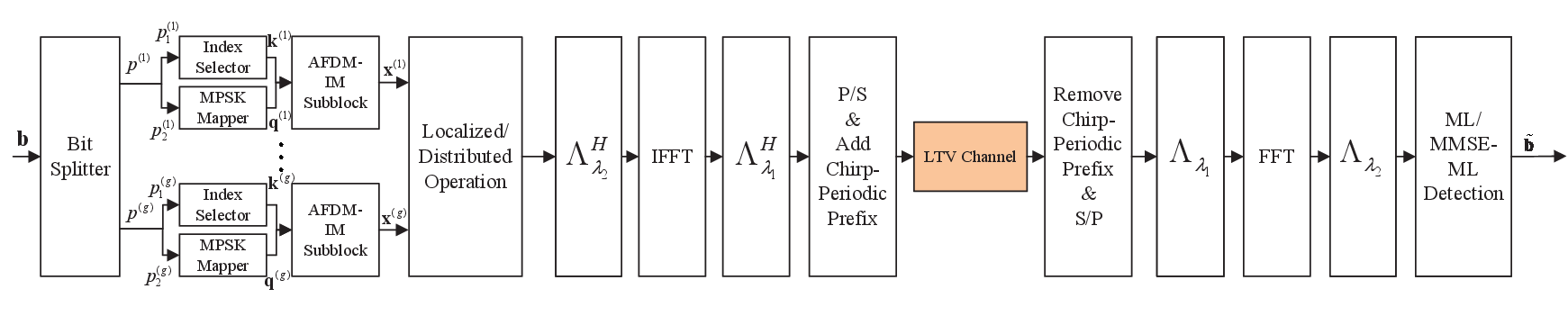}}
	\vspace{-0.6cm}
	\caption{{Transceiver structure of the proposed AFDM-IM scheme.}}
	\label{system-model}  
	\vspace{-4mm}
\end{figure*}

Nevertheless, there has been a lack of investigation on how to boost the data rates of the AFDM system. Index modulation (IM) techniques seem to be a promising solution~\cite{7929332}. Different from conventional modulation schemes, the IM exploits the state of the transmission resource (e.g., inactivation/activation of the antenna, and the position of the pulse) to carry extra information bits~\cite{7509396,8004416,9761226}.
This type of transmission technique boosts data rates without consuming additional energy, well-matched with the concept of low-power energy efficient systems in 5G.
In addition, for the subcarrier-index modulation, reducing the number of activated subcarriers can also lower the ICI over the LTV channel~\cite{8703169}.

Against the background, we propose an IM-assisted AFDM scheme, named AFDM-IM.
Specifically, we introduce the IM concept into the AFDM framework to achieve higher data rates.
In the proposed AFDM-IM scheme, the information bits are carried by both the modulation symbols as well as the activation states of the subsymbols in DAF domain.
Different from the OFDM-IM frameworks, the AFDM-IM scheme can exploit multipath diversity to achieve better performance.
We investigate the bit error rate (BER) performance of the proposed AFDM-IM scheme with and without satisfying the full diversity condition of AFDM.
Besides, to reduce the complexity of IM detection, we propose two grouping strategies, namely as localized and distributed strategies,
{and explore the performance of these two strategies.}
We develop both the maximum-likelihood (ML) and the joint minimum mean square error (MMSE)-ML detectors to recover the information bits.
Moreover, we derive an upper bound on the average BER of the proposed AFDM-IM scheme by using ML detection with perfect and imperfect channel state information (CSI).
Simulation results verify the tightness of the derived upper bound in the high signal-to-noise ratio (SNR) region. Last but not least, we compare the performance of the proposed AFDM-IM scheme with those of AFDM, classical OFDM, and OFDM-IM schemes.
The results show the superior performance of the proposed AFDM-IM scheme. Besides, the index bits can achieve higher-diversity-order protection than the modulated bits even when the full diversity condition of AFDM is not satisfied.

$Notations$: The notations $(\cdot)^{T}$, ${(\cdot)^{H}}$ and $(\cdot)^{-1}$ denote the transpose and conjugate transpose and inversion operations, respectively. $\left\lfloor \cdot  \right\rfloor $ represents the floor function.
$C(n,m)$ denotes the combination, which means that $m$ elements are randomly selected from $n$ elements.
$[\cdot]_{N}$ and $\left\| {\cdot} \right\|$ are modulo $N$ and Euclidean norm, respectively.
$\rm diag(\cdot)$ transforms a vector into a diagonal matrix. $\mathbb{C}^{M \times N}$ and ${\bf{I}}_{N}$ are an $M\times N$ matrix with complex entries and an $N\times N$ identity matrix, respectively. ${\rm rank}(\cdot)$ represents the rank of a matrix.
\vspace{-2mm}

\section{System Model of AFDM-IM}
\subsection{Transmitter}
Fig.~\ref{system-model} exhibits the transceiver structure of the proposed AFDM-IM scheme. The IM is performed in the DAF domain.
Each transmitted frame consists of $N$ DAF domain subsymbols.
To perform IM, the transmitted bits stream ${{\bf{b}}}$ is divided into $g$-groups by bit splitter, and the number of subsymbols in each group is $n=N/g$. Taking the $i$-${\rm th}$ ($i=1,\ldots,g$) group as an example, ${p}^{(i)}$ bits are first divided into ${p}^{(i)}_{1}$ index bits and ${p}^{(i)}_{2}$ modulated bits. Then, the index selector randomly selects $m$ indices from $n$ indices to activate based on the input ${p}^{(i)}_{1}$ index bits, and the remaining $n-m$ indices are inactive.
The output of the index selector can be represented as ${\bf k}^{(i)}=\{{k}^{(i)}_{1},\ldots,{k}^{(i)}_{m}\}$, where ${k}^{(i)}_{j}\in[1,\ldots,n]$, and $j\in[1,\ldots,m]$. Meanwhile, the ${p}^{(i)}_{2}$ modulated bits are fed to the $M$-PSK mapper, generating ${\bf q}^{(i)}=\{{q}^{(i)}_{1},\ldots,{q}^{(i)}_{m}\}$. From ${\bf k}^{(i)}$ and ${\bf q}^{(i)}$, we can obtain the $i$-$\rm{th}$ sub-symbol ${\bf x}^{(i)}\in{\mathbb{C}}^{n\times1}$.
For more clarity, all possible forms of the $i$-th sub-symbol ${\bf x}^{i}$ are given in Table~\ref{table1}, where $n=4$, and $m=1$.
 Therefore, the number of index bits ${p}^{(i)}_{1}={\rm log}_2\left\lfloor C(n,m) \right\rfloor$, and the number of modulated bits ${p}^{(i)}_{2}=m{\rm log}_2(M)$.
We assume that each group has the same $m$. Hence, the total number of information bits carried by the transmitted signal per frame can be calculated as $p=g{\rm log}_2\left\lfloor C(n,m) \right\rfloor+gm{\rm log}_2(M)$.
\begin{figure}[t]
	\center
	\includegraphics[width=2.5in,height=1.1in]{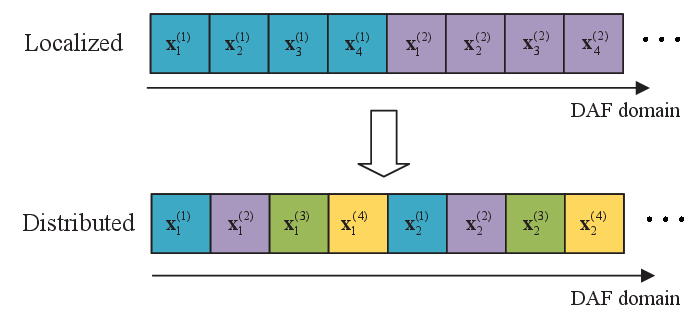}
	\vspace{-0.3cm}
	\caption{{An example of localized and distributed strategies, where $N=16$ and $g=4$.}}
	\label{centralised-distributed}  
	\vspace{-4mm}
\end{figure}
\begin{table}[t]
\caption{{An Example of the Subblock Activation State in the Proposed AFDM-IM Schemes, where $n=4$, and $m=1$. }}\label{table1}
\centering
\renewcommand\arraystretch{1.4}
\setlength{\tabcolsep}{6mm}{
\begin{tabular}{|c|c|}
\hline
Index Bits & ${\bf x}^{(i)}$ \\ \hline \hline
{[0~0]}  & {$[q^{(i)}_{1},~0,~0,~0]^T$}           \\ \hline
{[0~1]}  & {$[0,~q^{(i)}_{1},~0,~0]^T$}           \\ \hline
{[1~0]}  & {$[0,~0,~q^{(i)}_{1},~0]^T$}           \\ \hline
{[1~1]}  & {$[0,~0,~0,~q^{(i)}_{1}]^T$}           \\ \hline
\end{tabular}}
\vspace{-4mm}
\end{table}

After the above operations, the transmitted symbol ${{\bf x}=[{\bf x}^{(1)};\ldots;{\bf x}^{(g)}]}\in{\mathbb{C}}^{N\times1}$ in the DAF domain can be generated.
Then, we need to select the subsymbol grouping strategy. As shown in Fig.~\ref{centralised-distributed}, 
we offer two grouping strategies, i.e., localized and vector distributed ones~\cite{7330022}.
The localized strategy groups subsymbols that are continuous in DAF domain, i.e., the subsymbols of each group are adjacent.
In contrast, for the distributed strategy, the subsymbols of each group are uniformly distributed over the entire DAF domain, and the subsymbols of each group have the same spacing in the DAF domain.
Subsequently, the inverse DAFT is applied to convert ${\bf x}$ into a time domain signal ${\bf s}$,
\begin{equation}\label{1}
  {\bf s}=\Lambda _{{{\lambda}_{1}}}^{H}{{\bf F}^{H}}\Lambda _{{{\lambda}_{2}}}^{H}{\bf x},
\end{equation}
where the matrix $\Lambda _{{{\lambda}}}={\rm diag}(e^{-j2\pi {\lambda}{u^2}})$, $u=0,\ldots,N-1$, ${{\bf F}\in{\mathbb{C}}^{N\times N}}$ is discrete Fourier transform matrix with ${\bf F}[u,\bar{u}]=\frac{1}{\sqrt{N}}{e^{-j2\pi {\bar{u}}{u}/N}}$, and $u,\bar{u}=0,1,\ldots,N-1$. The form of \eqref{1} can also be expressed as
\begin{equation}\label{2}
{s}(t)=\frac{1}{\sqrt{N}}
\sum\limits_{u=0}^{N-1}{{{\bf x}{[u]}}}{{e}^{j2\pi \left( {{\lambda }_{2}}{{u}^{2}}+\frac{u}{NT}t+\frac{{{\lambda }_{1}}}{{{T}^{2}}}{{t}^2} \right)}},
\end{equation}
where $T$ is the sampling interval, and ${T}_{p}=NT$ represents the symbol period.
Similarly, to avoid inter-symbol interference (ISI), a chirp periodic prefix (CPP) needs to be added to the transmit signal. Without loss of generality, we assume that the length of CPP is greater than the maximum path delay spread. Therefore, ISI can be ignored. At last, the transmit signal of the AFDM-IM scheme is sent to the receiver.

\subsection{Receiver}
Without loss of generality, the LTV channel is considered in this paper, which can be modeled as
\begin{equation}\label{3}
h(\tau ,\nu )=\sum\limits_{i=1}^{P}{{{h}_{i}}\delta (\tau -{{\tau }_{i}})}\delta (\nu -{{\nu }_{i}}),
\end{equation}
where $\delta (\cdot )$ is the unit pulse function, $P$ denotes the number of fading paths, and ${h}_{i}\sim\mathcal{CN}(0,1/P)$, ${\tau }_{i}$ and ${\nu }_{i}$ represent the channel coefficient, delay and Doppler of the $i$-$\rm th$ path, respectively.
It is worth mentioning that this channel model is a generalized model that allows each delay tap to have different Doppler shift values, i.e., $i,j=1,\ldots,P$, ${\tau}_{i}={\tau}_{j}$, and ${\nu}_{i}\ne{\nu}_{j}$~\cite{9473655}.
Then, after removing the CPP, the received signal in the time domain can be written as
\begin{equation}\label{4}
r(t)=\sum\limits_{i=1}^{P}{{{h}_{i}}s(t-{{\tau }_{i}}){{e}^{j2\pi {{\nu }_{i}}t}}}+w(t),
\end{equation}
where $w(t)\sim\mathcal{CN}(0,N_0)$ is the complex additive white Gaussian noise (AWGN). Then, one can obtain the $\bar u$-$\rm th$ received symbol in the DAF domain as
\begin{align}\label{5}
   {{{\bf y}}{{{[\bar u]}}}}=&\frac{1}{NT}\int_{0}^{NT}{r(t){{e}^{-j2\pi \left( {{\lambda }_{2}}{{{\bar{u}}}^{2}}+\frac{{\bar{u}}}{NT}t+\frac{{{\lambda }_{1}}}{{{T}^{2}}}{{t}^{2}} \right)}}}dt \nonumber\\
  =&\frac{1}{NT}\sum\limits_{i=1}^{P}{{{h}_{i}}}\sum\limits_{u=0}^{N-1}{{{\bf x}{[u]}}{{e}^{j2\pi \left[ {{\lambda }_{2}}({{u}^{2}}-{{{\bar{u}}}^{2}})-\frac{u{{\tau }_{i}}}{NT}+\frac{{{\lambda }_{1}}\tau _{i}^{2}}{{{T}^{2}}}
   \right]}}} \nonumber\\
 & \times \int_{0}^{NT}{{{e}^{j\frac{2\pi }{NT}\left[ u-\left( \bar{u}-NT{{\nu }_{i}}+2\frac{{{\lambda }_{1}}}{T}N{{\tau }_{i}} \right) \right]t}}}dt+{{{\overline{\bf w}}}{[\bar u]}},
\end{align}
where ${{{\overline{\bf w}}}}$ is the filtered output of ${{w}(t)}$. Let us define ${\alpha}_{i}=NT{\nu_{i}}$, ${\alpha}_{i}\in[-{\alpha}_{\rm max},{\alpha}_{\rm max}]$, and ${l}_{i}={\tau}_{i}/T$, ${l}_{i}\in[0,{l}_{\rm max}]$.
To simplify the analysis, we consider $\alpha_i$ to be an integer Doppler shift value.
Hence, \eqref{5} can be rewritten as
   \begin{align}\label{6}
   {{{\mathbf{y}}}{{[{\bar u}]}}}=&\sum\limits_{i=1}^{P}{{{h}_{i}}}\sum\limits_{u=0}^{N-1}{{{\mathbf{x}}{[u]}}{{e}^{j\frac{2\pi }{N}\left[ N{{\lambda }_{2}}({{u}^{2}}-{{{\bar{u}}}^{2}})-u{{l}_{i}}+N{{\lambda }_{1}}l_{i}^{2} \right]}}} \nonumber\\
 & \times \delta \left( {{[u-\left( \bar{u}-{{\alpha }_{i}}+2N{{\lambda }_{1}}{{l}_{i}} \right)]}_{N}} \right)+{{{\overline{\bf w}}}{[\bar u]}}.
\end{align}
For the proposed AFDM-IM scheme, when setting ${\lambda}_{1}=(2\alpha_{\rm max}+1)/{2N}$, ${\lambda}_{2}$ is an irrational number, and the maximum possible number of paths ${P}_{\rm max}=({{l}_{\rm max}+1})(2{\alpha}_{\rm max}+1)$ is less than the number of subsymbols $N$, the modulated bits are able to obtain full diversity~\cite{9473655}.
According to~\eqref{6}, we can obtain the effective channel matrix ${\bf H}_{\rm eff}=\sum\nolimits_{i=1}^{P}{h}_{i}{\bf H}_{i}$, and ${\bf H}_{i}$ can be expressed as
\begin{equation}\label{7}
  {{\bf H}_{i}}[\bar{u},{u}]\!=\!\left\{ \begin{array}{*{35}{l}}
   \!{{e}^{j\frac{2\pi }{N}\left[ N{{\lambda }_{2}}({{u}^{2}}-{{{\bar{u}}}^{2}})-u{{l}_{i}}+N{{\lambda }_{1}}l_{i}^{2} \right]}}, \!&u={{\left[ \bar{u}+loc_i \right]}_{N}}  \\
   \!0, \!& {\rm otherwise},
\end{array} \right.
\end{equation}
where $loc_i=-{{\alpha }_{i}}+2N{{\lambda }_{1}}{{l}_{i}}$. This means that each row of the ${\bf H}_{\rm eff}$ matrix has $P$ non-zero elements. Thus, the AFDM scheme can achieve a full diversity gain.
It is worth noting that in practical communication systems, perfect CSI might be difficult to obtain due to the limitations of channel estimation algorithms and low-resolution quantization channel feedback.
According to~\cite{6094132}, imperfect CSI can be modeled as{\footnote{We do not consider the delay and Doppler estimation errors in this paper to facilitate the derivation of the theoretical BER upper bound. In the future, we will investigate the impact of delay and Doppler estimation errors on the performance of the proposed AFDM-IM scheme.}}
\begin{equation}\label{11}
{{\bar{h}}_{i}}=\sqrt{1-{{\rho }^{2}}}{{h}_{i}}+\rho \phi,
\end{equation}
where $\phi\sim\mathcal{CN}(0,1)$, $\phi$ is independent for ${{h}_{i}}$, and ${\rho\in[0,1]}$. When $\rho=0$, \eqref{11} can be seen as the case of perfect CSI.

The AFDM scheme multiplexes symbols over $N$ subsymbols in the DAF domain.
There will be interference between $g$ AFDM-IM groups. Therefore, we first introduce the MMSE equalisation, which can be given by
\begin{equation}\label{8}
{{\widehat{\bf x}}_{\rm MMSE}}={\overline{\bf H}}_{\rm eff}^{H}{{\left( {{\overline{\bf H}}_{\rm eff}}{\overline{\bf H}}_{\rm eff}^{H}+{\frac{1}{\gamma}}{{\bf I}_{N}} \right)}^{-1}}{\bf y},
\end{equation}
where ${\overline{\bf H}_{\rm eff}}=\sum\nolimits_{i=1}^{P}{\bar{h}}_{i}{\bf H}_{i}$ represents the imperfect ${{\bf H}_{\rm eff}}$, and $\gamma ={{E}_{b}}/{{N}_{0}}$ denotes the average SNR. Without loss of generality, we ignore the energy of the CPP, and ${{E}_{b}}$ can be calculated as ${{E}_{b}}=gm/p$.
For the $u$-$\rm th$ symbol in the AFDM-IM scheme, its estimation can be given by
\begin{equation}\label{8.1}
\widehat{x}_{{\rm MMSE},u}={\vec f}_{u}^{H}{{\Re }^{-1}}{\mathbf{y}},
\end{equation}
where $\vec{f}_{u}\in{\mathbb{C}}^{N\times 1}$ represents the $u$-{\rm th} column of ${\overline{\bf H}}_{\rm eff}$, and ${\Re }={{{\overline{\bf H}}_{\rm eff}}{\overline{\bf H}}_{\rm eff}^{H}+{\frac{1}{\gamma}}{{\bf I}_{N}} }$.
Then, the equalised signals are divided into $g$ groups, and each group is independently subjected to the ML detection. For the $i$-$\rm th$ group, the ML detection can be given by
\begin{equation}\label{9}
\left( {{{\tilde{\bf k}}}^{(i)}},{{{\tilde{\bf q}}}^{(i)}} \right)=\underset{{{{{\bf k}}}^{(i)}},{{{{\bf q}}}^{(i)}} }{\rm{arg~min}}\,{{\left\| {{{\widehat{\bf x}}}^{(i)}_{\rm MMSE}}-{{\bf R}^{(i)}}{{\bf x}^{(i)}} \right\|}^{2}},
\end{equation}
where ${{\bf x}^{(i)}}$ contains all the possibilities of the $i$-$\rm th$ group in the AFDM-IM scheme, ${{\mathbf{R}}^{(i)}}={\rm diag}({R}^{(i)}_{1},\ldots,{R}^{(i)}_{l},\ldots,{R}^{(i)}_{n})$,
%
${R}^{(i)}_{l}{{{x}}^{(i)}_{l}}$ represents the mean of $\widehat{x}^{(i)}_{{\rm MMSE},l}$,
and each element of ${{\mathbf{R}}^{(i)}}$ can be calculated as  ${R}^{(i)}_{l}={\vec f}_{l}^{H}{{\Re }^{-1}}{\vec f}_{l}$.
Alternatively, we can use the ML detection directly to achieve the optimal BER performance, though the computational complexity of the ML detection grows exponentially as $N$ increases.
The ML detection can be formulated as
\begin{equation}\label{10}
\left(\widetilde{\mathbf{k}},\widetilde{\mathbf{q}} \right)= \underset{\mathbf{k},\mathbf{q}}{\rm{arg~min }}\,{{\left\| {\bf y}-{{\overline{\bf H}}_{\rm eff}}{{{\mathbf{x}}}} \right\|}^{2}}.
\end{equation}
Finally, after the index symbols and modulated symbols are estimated by the MMSE-ML or ML detection algorithms described above, the information bits stream ${\widetilde{\bf b}}$ can be easily recovered by the index detector and $M$-PSK demapper.
\vspace{-2mm}

\section{Performance Analysis For AFDM-IM}
In this section, we analyse the average bit error probability (ABEP) of the proposed AFDM-IM scheme by using the ML detection in~\eqref{10}.
To facilitate the elaboration, we can write the received signal in the DAF domain as
\begin{equation}\label{12}
 {\bf y}=\sum\limits_{i=1}^{P}{{{h}_{i}}{{\bf H}_{i}}{\bf x}+\overline{\bf w}=\Upsilon_{\bf x}}{\bf h}+\overline{\bf w},
\end{equation}
where ${\bf h}=[{h}_{1},\ldots,{h}_{P}]\in{\mathbb{C}}^{P\times1}$, and $\Upsilon_{\bf x}=[{{\bf H}_{1}{\bf x}},{{\bf H}_{2}{\bf x}},\ldots,{{\bf H}_{P}{\bf x}}]\in{{\mathbb{C}}^{N\times P}}$.
Thus, in the case of a channel estimation error as shown in~\eqref{11}, the conditional pairwise error probability (PEP) between the transmitted symbol ${\bf x}_{i}$ and the estimated symbol ${\bf x}_{j}$ can be expressed as~\cite{6587554}
\begin{align}\label{13}
  & {\rm Pr}\left({{\bf x}_{i}}\to {{\bf x}_{j}}|\bar{\bf h} \right) \nonumber\\
 &\!=\!Q\left( \frac{{{\left\| \left( {{\Upsilon }_{{{\bf x}_{i}}}}\!\!-\!\!{{\Upsilon }_{{{\bf x}_{j}}}} \right)\!\bar{\bf h} \right\|}^{2}}}{\sqrt{2\rho {{\left\| \Upsilon _{{{\bf x}_{i}}}^{H}\left( {{\Upsilon }_{{{\bf x}_{i}}}}\!\!-\!\!{{\Upsilon }_{{{\bf x}_{j}}}} \right)\!\bar{\bf h} \right\|}^{2}}\!\!\!+\!\!2{{N}_{0}}{{\left\| \left( {{\Upsilon }_{{{\bf x}_{i}}}}\!\!-\!\!{{\Upsilon }_{{{\bf x}_{j}}}} \right)\!\bar{\bf h} \right\|}^{2}}}} \right),
\end{align}
where $Q\{\cdot\}$ is the tail distribution function of the standard Gaussian distribution. Further, we can rewrite
${{\left\| \Upsilon _{{{\bf x}_{i}}}^{H}\left( {{\Upsilon }_{{{\bf x}_{i}}}}-{{\Upsilon }_{{{\bf x}_{j}}}} \right)\bar{\bf h} \right\|}^{2}}={{\left\| {{\Upsilon }_{{{\bf x}_{i}}}} \right\|}^{2}}{{\left\| \left( {{\Upsilon }_{{{\bf x}_{i}}}}-{{\Upsilon }_{{{\bf x}_{j}}}} \right)\bar{\bf h} \right\|}^{2}}$. Therefore, \eqref{13} can be further expressed as
\begin{equation}\label{14}
  {\rm Pr}\left( {{\bf x}_{i}}\to {{\bf x}_{j}}|\bar{\bf h} \right)=Q\left( \sqrt{\frac{{{\left\| \left( {{\Upsilon }_{{{\bf x}_{i}}}}-{{\Upsilon }_{{{\bf x}_{j}}}} \right)\bar{\bf h} \right\|}^{2}}}{2\rho{{\left\| {{\Upsilon }_{{{\bf x}_{i}}}} \right\|}^{2}} +2{{N}_{0}}}} \right). \\
\end{equation}
According to~\cite{10159363}, we can approximate
\begin{equation}\label{15}
Q(x)\cong \frac{1}{12}{{e}^{-{{x}^{2}}/2}}+\frac{1}{4}{{e}^{-2{{x}^{2}}/3}}.
\end{equation}
Then, the unconditional PEP can be given by
\begin{equation}\label{16}
{\rm Pr}\left( {{\bf x}_{i}}\to {{\bf x}_{j}} \right)\cong {{E}_{{\bar{\bf h}}}}\left\{ \frac{1}{12}{{e}^{-{{q}_{1}}\Phi }}+\frac{1}{4}{{e}^{-{{q}_{2}}\Phi }} \right\},
\end{equation}
where $q_1=1/(4{N_0}+4\rho{{\left\| {{\Upsilon }_{{{\bf x}_{i}}}} \right\|}^{2}})$, $q_2=1/(3{N_0}+3\rho{{\left\| {{\Upsilon }_{{{\bf x}_{i}}}} \right\|}^{2}})$, and $\Phi={{{\left\| \left( {{\Upsilon }_{{{\bf x}_{i}}}}-{{\Upsilon }_{{{\bf x}_{j}}}} \right)\bar{\bf h} \right\|}^{2}}}$.
Let us define ${\Psi}=E\{{\bar{\bf{h}}}{\bar{\bf{h}}}^{H}\}$.
Thus, the probability density function of ${\bar{\bf{h}}}$ can be expressed as
\begin{equation}\label{17}
f({ \bar{\bf h}})=\frac{{{\pi }^{-{\rm rank}(\Psi)}}}{\det (\Psi )}\exp (-{{\bar{\bf h}}^{H}}{\Psi}^{-1} {\bar{\bf h}}).
\end{equation}
Without loss of generality, it is assumed that the proposed AFDM-IM scheme satisfies the full diversity condition, so that ${\rm rank}(\Psi)=P$.
Based on~\eqref{17}, the unconditional PEP in~\eqref{16} can be calculated as
\begin{align}\label{18}
{\rm Pr}\left( {{\bf x}_{i}}\to {{\bf x}_{j}} \right)& \cong \frac{1/12}{{\rm det}\left({{\bf I}_{N}}+{{q}_{1}\Psi{\bf A}}\right)}+\frac{1/4}{{\rm det}\left({{\bf I}_{N}}+{{q}_{2}\Psi{\bf A}}\right)}\nonumber\\
& \approx \frac{1}{12}\prod\limits_{i=1}^{P}{\frac{1}{1+\frac{{{q}_{1}}\kappa _{i}}{P}}}+\frac{1}{4}\prod\limits_{i=1}^{P}{\frac{1}{1+\frac{{{q}_{2}}\kappa _{i}}{P}}},
\end{align}
where ${\bf A}=\left({{\Upsilon }_{{{\bf x}_{i}}}}-{{\Upsilon }_{{{\bf x}_{j}}}}\right)\left({{\Upsilon }_{{{\bf x}_{i}}}}-{{\Upsilon }_{{{\bf x}_{j}}}}\right)^{H}$, and $\kappa _{i}$ is the eigenvalue of matrix ${\bf A}$. In the high SNR region,~\eqref{18} can be approximated as
\begin{equation}\label{19}
{\rm Pr}\left( {{\mathbf{x}}_{i}}\to {{\mathbf{x}}_{j}} \right)\approx \frac{1/12}{\prod\limits_{i=1}^{P}{\frac{{{q}_{1}}\kappa _{i}}{P}}}+\frac{1/4}{\prod\limits_{i=1}^{P}{\frac{{{q}_{2}}\kappa _{i}}{P}}}.
\end{equation}
Finally, based on the obtained PEP, we can calculate the ABEP upper bound of the proposed AFDM-IM scheme with imperfect CSI as
\begin{equation}\label{20}
{{\rm Pr}_{\rm ABEP}}\approx\frac{1}{{{2}^{p}}p}\sum\limits_{{{\bf x}_{i}}}{\sum\limits_{{{\bf x}_{j}}}{{\rm Pr}\left( {{\bf x}_{i}}-{{\bf x}_{j}} \right)N\left( {{\bf x}_{i}},{{\bf x}_{j}} \right)}},
\end{equation}
where $N\left( {{\bf x}_{i}},{{\bf x}_{j}} \right)$ denotes the number of error bits for ${{\bf x}_{i}}$ when it is estimated as ${{\bf x}_{j}}$.

\begin{figure}[t]
	\center
	\includegraphics[width=2.1in,height=1.7in]{{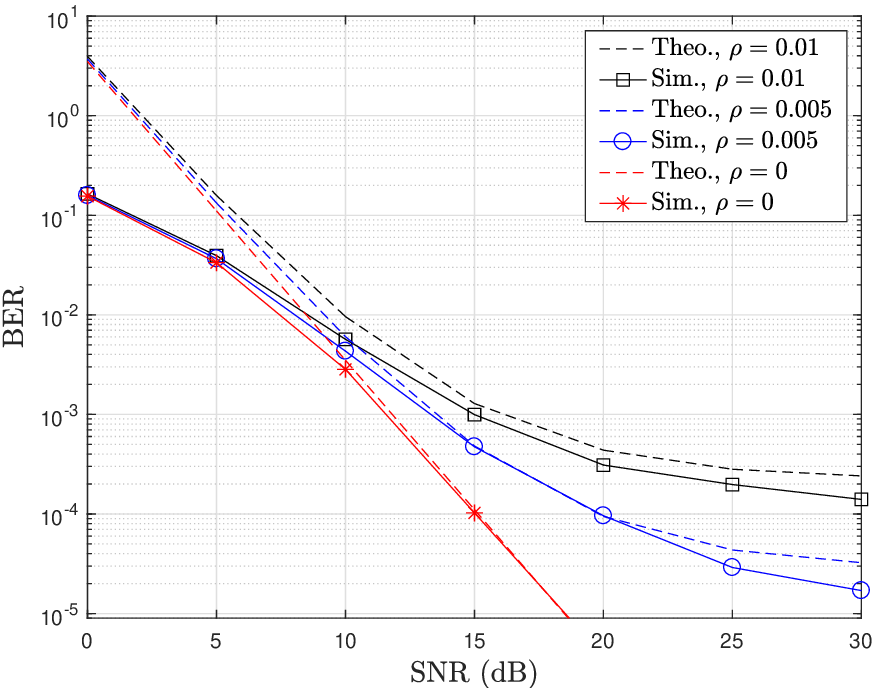}}
	\vspace{-0.2cm}
	\caption{Simulated and theoretical BER performance of the proposed AFDM-IM scheme by using the ML
detection with QPSK over an LTV channel, where $n=4$, $m=1$, $g=1$, $\rho=0/0.01/0.005$, $P=3$, ${l}_{\rm max}=0$, and ${\alpha_{\rm max}=1}$.}
	\label{Sim-Theo}  
	\vspace{-5mm}
\end{figure}

\section{Simulation Results and Discussions}
In this section, to verify the accuracy of the theoretical analysis, and evaluate the BER performance of the proposed AFDM-IM scheme, various Monte Carlo experiments are carried out over different channel parameters.

\begin{figure}[t]
	\center
	\includegraphics[width=2.4in,height=1.9in]{{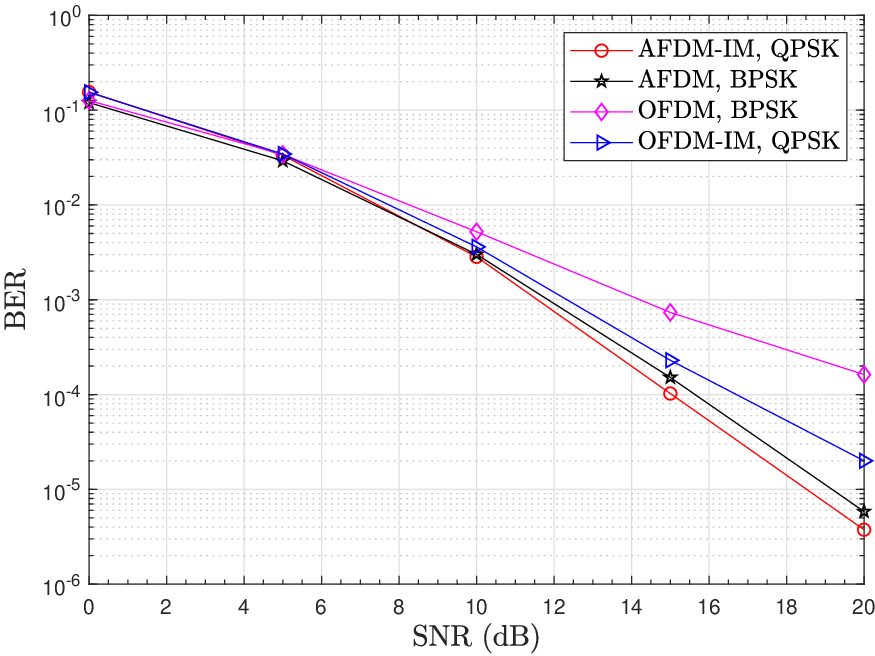}}
	\vspace{-0.2cm}
	\caption{BER performance of the proposed AFDM-IM scheme, conventional AFDM scheme, conventional OFDM scheme and OFDM-IM scheme by using the ML detection over an LTV channel, where $n=4$, $m=1$, $g=1$, $\rho=0$, $P=3$, ${l}_{\rm max}=0$, and ${\alpha_{\rm max}=1}$.}
	\label{BER-full-d}  
	\vspace{-4mm}
\end{figure}
\begin{figure}[t]
	\center
	\includegraphics[width=2.4in,height=1.9in]{{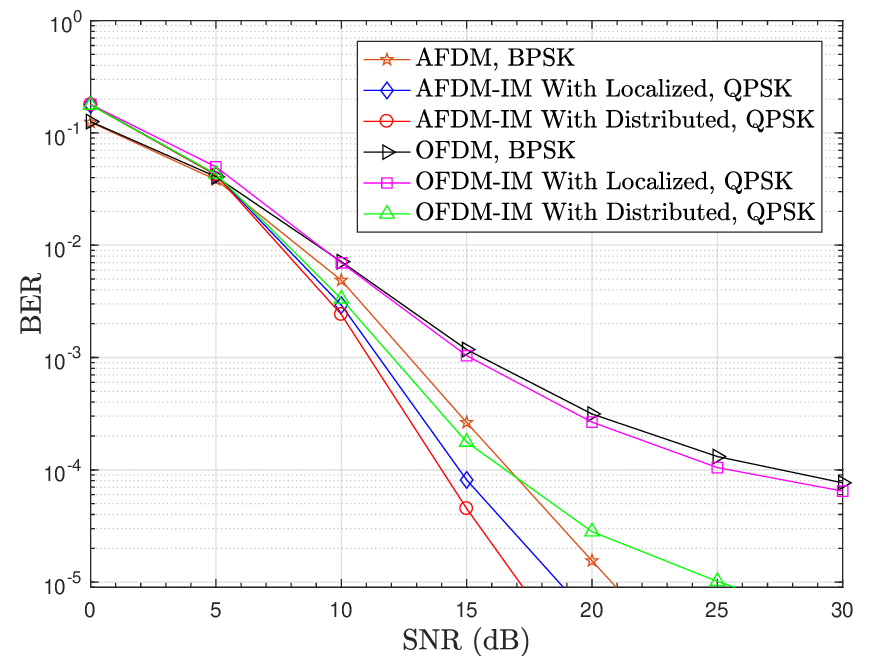}}
	\vspace{-0.2cm}
	\caption{BER performance of the proposed AFDM-IM scheme, conventional AFDM scheme, conventional OFDM scheme and OFDM-IM scheme by using the MMSE-ML/MMSE detection over an LTV channel, where $N=64$, $m=1$, $g=16$, $\rho=0$, $P=21$, ${l}_{\rm max}=2$, and ${\alpha_{\rm max}=3}$.}
	\label{BER-MMSE-V2}  
	\vspace{-5mm}
\end{figure}

\begin{figure}[t]
\vspace{-0.0cm}
\centering
\subfigcapskip=+2pt
\subfigure[\hspace{-0.2cm}]{\label{full-diversity}
\includegraphics[width=2.4in,height=1.9in]{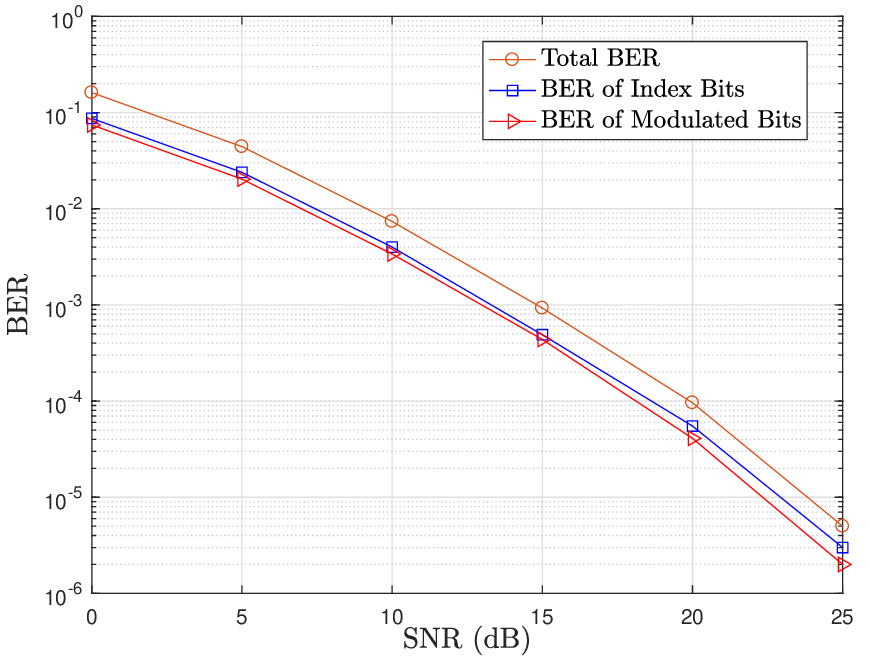}}
\vspace{+0.0cm}
\subfigure[\hspace{-0.2cm}]{\label{unfull-diversity}
\includegraphics[width=2.4in,height=1.9in]{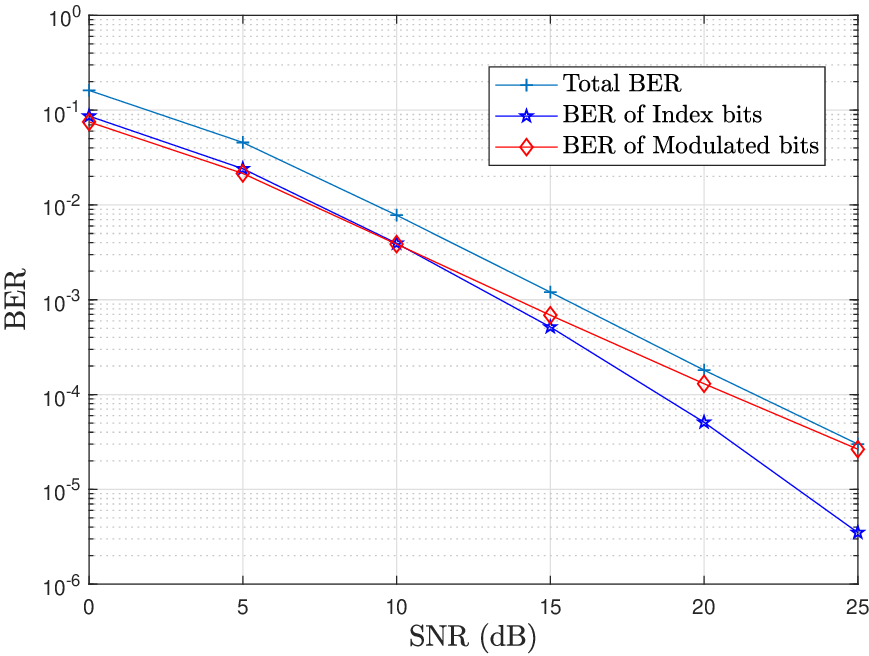}}
\vspace{-0.1cm}
\caption{BER performance of the index bits, modulated bits, and total bits for the proposed AFDM-IM scheme by using the ML
detection with QPSK over an LTV channel, where $n=4$, $m=1$, $g=1$, $P=2$, (a) ${l}_{\rm max}=0$, ${\alpha_{\rm max}=1}$, and (b) ${l}_{\rm max}=1$, ${\alpha_{\rm max}=1}$.}\label{unfull-full}
\vspace{-0.4cm}
\end{figure}
Fig.~\ref{Sim-Theo} illustrates the theoretical and simulated BER results of the proposed AFDM-IM scheme with different channel estimation errors. One can easily observed that in the perfect CSI case (i.e., $\rho=0$),  the theoretical and simulated curves match well in the high SNR region. The results indicated the validity of the theoretical derivation in~\eqref{20}.
As the channel estimation error (i.e., $\rho$) increases, 
the BER performance of the proposed AFDM-IM scheme deteriorates, and the error floor occurs.

The performance comparison between the proposed AFDM-IM scheme and the conventional AFDM scheme is illustrated in Fig.~\ref{BER-full-d}, where the performance of conventional OFDM and OFDM-IM schemes are also provided as benchmarks.
To ensure the fairness of the comparison, the spectrum efficiencies of both the proposed AFDM-IM scheme and the benchmark schemes are set to 1. One can observe that the proposed AFDM-IM scheme achieves superior performance over the benchmark schemes.
For example, at a BER level of ${10}^{-4}$, the proposed AFDM-IM scheme yields gains of about 5 $\rm dB$ and 2 $\rm dB$ compared to OFDM and OFDM-IM schemes over an LTV channel, respectively. Besides, the proposed AFDM-IM scheme outperforms the AFDM scheme at a BER level of $10^{-5}$, and achieves a gain of about 1 $\rm dB$.
Both the proposed AFDM-IM scheme and AFDM scheme outperform the OFDM schemes, which shows that AFDM is a very promising technique for the LTV~channel.

Fig.~\ref{BER-MMSE-V2} exhibits the BER performance of the proposed AFDM-IM scheme with localized/distributed strategies and other benchmark schemes by using the MMSE-ML detection.
Similar to the results in Fig.~\ref{BER-full-d}, the proposed AFDM-IM scheme with localized/distributed strategies can achieve better BER performance than the AFDM, OFDM and OFDM-IM schemes.
For example, the proposed AFDM-IM scheme with distributed strategy shows a gain of about 4 $\rm dB$ compared to the AFDM scheme.
Moreover, we can find that the distributed AFDM-IM scheme exhibits better BER performance than the localized AFDM-IM scheme.
This is because 
after MMSE equalization of the received signal, the mean of each estimated subsymbol in DAF domain is different.
Compared to the localized strategy, using the distributed strategy allows for a greater difference in the mean values of the subsymbols within the group in the DAF domain, resulting in higher coding gain from IM.
For the OFDM schemes, the OFDM-IM with distributed strategy achieves significant performance gains over the OFDM-IM scheme with localized strategy.
This is expected because the distributed strategy ensures the channel coefficients
on subcarriers within a subcarrier group are approximately
statistically independent. The independence of channel fading makes the different subcarrier activation states easier to be differentiated, which in turn improves the coding gain from index modulation~\cite{7330022}.
Furthermore, thanks to the identity that the channel taps do not overlap in the effective channel matrix, the AFDM scheme harvests the multipath diversity, outperforming OFDM~schemes.



Further, in Fig.~\ref{unfull-full}, we investigate the BER performance of the proposed AFDM-IM scheme when the maximum possible number of paths ${P}_{\rm max}$ is greater than and less than the number of subsymbols $N$, respectively.
To facilitate elaboration, in Fig.~\ref{unfull-full}, we give the BER of index bits, BER of modulated bits, and total BER of the proposed AFDM-IM scheme.
From Fig.~\ref{full-diversity} (i.e., ${P}_{\rm max}<N$), we can observe that the BER performance of index bits and modulated bits are similar as both can achieve full diversity.
In contrast, as shown in Fig.~\ref{unfull-diversity} (i.e., ${P}_{\rm max}>N$), the BER performance of modulated bits deteriorates while that of index bits is not affected.
This is because when ${P}_{\rm max}>N$, the channel matrix does not have enough space to separate the channel taps, which will lead to overlapping between different paths and the full diversity condition of AFDM is not satisfied as shown in~\cite{9473655}.
Thus, the modulated bits in the AFDM-IM scheme is unable to obtain full diversity gain when the channel paths are overlapped.
However, due to the IM, i.e., some information bits are transmitted through the activated state of the subsymbols in the DAF domain, the performance of index bits can still maintain the diversity gain.
Such a case usually occurs in large delay/Doppler or multi-antenna communication scenarios, in which the intelligent integration of IM techniques and AFDM schemes is of great significance in achieving highly reliable transmission.

%

\begin{figure}[t]
	\center
	\includegraphics[width=2.2in,height=1.8in]{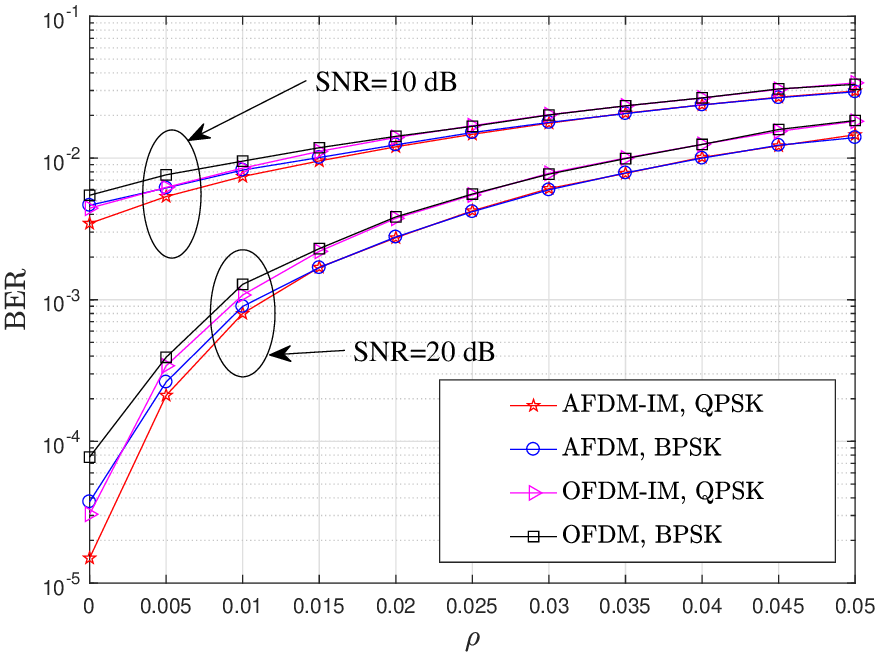}
	\vspace{-0.2cm}
	\caption{BER results of the proposed AFDM-IM scheme, conventional AFDM scheme, conventional OFDM scheme and OFDM-IM scheme by using ML detection over an LTV channel, where $n=4$, $m=1$, $g=1$, $P=3$, ${l}_{\rm max}=1$, and ${\alpha_{\rm max}=1}$.}
	\label{BER-imperfect}  
	\vspace{-5mm}
\end{figure}

Finally, Fig.~\ref{BER-imperfect} compares the BER performance of the proposed AFDM-IM, AFDM, OFDM and OFDM-IM schemes with imperfect CSI, when $P_{\rm max}>N$ (i.e., the full diversity condition of AFDM is not satisfied). One can notice that the proposed AFDM-IM scheme has a lower BER level than the AFDM, OFDM and OFDM-IM schemes with different accuracy of CSI. This shows that our proposed AFDM-IM scheme 
exhibits robustness against CSI uncertainty.
\vspace{-1mm}
\section{Conclusion}
In this paper, an index modulation scheme based on the AFDM framework has been developed. The proposed AFDM-IM scheme has higher data rate with respect to the AFDM scheme. We have derived theoretical BER upper bounds for the proposed AFDM-IM schemes with perfect and imperfect CSI cases.
Simulation results have been presented to verify the accuracy of the theoretical analysis for the proposed AFDM-IM scheme.
The performance comparison results with the benchmark schemes have shown that the proposed AFDM-IM scheme has significant performance improvement in the LTV channel.
Moreover, we have evaluated the BER performance of index bits and modulated bits for the proposed AFDM-IM scheme with and without satisfying the full diversity condition of AFDM.
The results have demonstrated that the index bits can have a stronger diversity protection than the modulated bits
when the full diversity condition of AFDM is not satisfied.
Thanks to the benefits mentioned above, the proposed AFDM-IM scheme is a promising technique for high-mobility communications.
In the future, we will investigate the AFDM-IM schemes that offer both high reliability and high data rates, and consider the effect of Doppler and delay estimation errors on the AFDM-IM systems.
\vspace{-2mm}

\bibliographystyle{IEEEtran}  

\end{document}